\documentclass[twocolumn,a4,prl,preprintnumbers,amsmath,amssymb]{revtex4}

\newcommand{\m}[1]
{\mathrm{#1}}

\newcommand{\be}{\begin{eqnarray}}

\newcommand{\ee}
{\end{eqnarray}}

\newcommand{\X}{\mathrm{X^0}}
\newcommand{\Xn}{\mathrm{X^{-}}}
\newcommand{\Xp}{\mathrm{X^{+}}}

\newcommand{\ket}[1]{\mbox{$| #1 \rangle$}}
\newcommand{\bra}[1]{\mbox{$\langle #1 |$}}

\usepackage{graphicx}
\usepackage{dcolumn}
\usepackage{bm}
\usepackage{times}
\usepackage{amsmath}
\usepackage{amsfonts}
\usepackage{amssymb}
\usepackage{color}
\definecolor{red}{rgb}{1,0,0}
\definecolor{blue}{rgb}{0,0,1}

\begin{document}

\title{Dynamic nuclear spin polarization in resonant laser spectroscopy of a quantum dot}
\author{A. H\"ogele$^{1}$, M. Kroner$^{2}$, C. Latta$^{2}$, M. Claassen$^{3}$, I. Carusotto$^{4}$, C. Bulutay$^{2,5}$,
and A. Imamoglu$^{2}$}

\affiliation{$^1$Fakult\"at f\"ur Physik and CeNS,
Ludwig-Maximilians-Universit\"at M\"unchen, D-80539 M\"unchen,
Germany}

\affiliation{$^2$Institute of Quantum Electronics, ETH Z\"urich,
CH-8093, Z\"urich, Switzerland}

\affiliation{$^3$Department of Applied Physics, Stanford
University, Stanford, California 94305, USA}

\affiliation{$^4$INO-CNR BEC Center and Dipartimento di Fisica,
Universit\`a di Trento, I-38123 Povo, Italy}

\affiliation{$^5$Department of Physics, Bilkent University,
Ankara, 06800, Turkey}

\date{\today}

\begin{abstract}
Resonant optical excitation of lowest-energy excitonic transitions
in self-assembled quantum dots lead to nuclear spin polarization
that is qualitatively different from the well known optical
orientation phenomena. By carrying out a comprehensive set of
experiments, we demonstrate that nuclear spin polarization manifests
itself in quantum dots subjected to finite external magnetic field
as locking of the higher energy Zeeman transition to the driving
laser field, as well as the avoidance of the resonance condition for
the lower energy Zeeman branch. We interpret our findings on the
basis of dynamic nuclear spin polarization originating from
non-collinear hyperfine interaction and find an excellent agreement
between the experimental results and the theoretical model.
\end{abstract}

\maketitle

The basic principles of optical nuclear spin orientation in solids
have been studied extensively in bulk semiconductors \cite{Meier}
and attracted a revived attention by recent optical studies of
semiconductor quantum dots (QDs). Dynamic nuclear spin
polarization (DNSP) in self-assembled InGaAs QDs has been reported
for quasi-resonant \cite{Lai2006} and non-resonant excitation
\cite{Eble2006,Tartakovskii2007}. On the basis of these
experiments and related theoretical studies, a comprehensive
picture of unidirectional optical orientation of QD nuclear spins
effected by light-polarization selective pumping was developed.
Early experiments carried out on positive and negative trions
\cite{Lai2006,Eble2006,Tartakovskii2007} as well as neutral
excitons \cite{Belhadj2009} had been used to demonstrate
bistability of DNSP as a function of magnetic field or incident
laser power and polarization
\cite{Eble2006,Tartakovskii2007,Maletinsky2007prb,Braun2006prb},
and to study nuclear spin buildup and decay dynamics
\cite{Maletinsky2007prl,Maletinsky2009}. In stark contrast to
non-resonant excitation however, bidirectional nuclear spin
orientation independent of photon polarization was observed in
resonant laser scattering of elementary transitions in neutral
\cite{Latta2009} and negatively charged QDs
\cite{Latta2009,Xu2009,Ladd2010}. A particularly striking feature
of resonant DNSP  using the higher energy Zeeman transition at
external magnetic fields exceeding $1$~T is the flat-top
absorption spectra, stemming from active locking of the QD
resonance to the laser frequency \cite{Latta2009,Xu2009}.
Remarkably, neutral and negatively charged QDs showed similar
spectral signatures in resonant spectroscopy despite substantially
different energy level diagrams: for both charge states, the
locking of the coupled electron-nuclear spin system to the
incident laser (dragging) was observed over tens of $\mu$eV
detunings to either side of the resonance \cite{Latta2009}.

In this Letter, we carry out a comprehensive experimental and
theoretical analysis of dragging in resonantly driven QD
transitions. We develop a microscopic model based on the effective
non-collinear hyperfine coupling that was first proposed by
Ref.~\cite{Huang2010} to explain nuclear spin relaxation in
self-assembled QDs. Our experiments demonstrate that the nature of
resonant DNSP depends drastically on whether the blue (higher
energy) or the red (lower energy) Zeeman transition is resonantly
excited; while the blue transition exhibits locking of the QD
resonance to the incident laser, nuclear spin polarization ensures
that the resonance condition is avoided for the red transition
\cite{Yang2011}. We also find that while the frequency range over
which blue Zeeman transition locking takes place varies from QD to
QD, the dependence of the corresponding dragging width on laser
power, scan speed and the magnetic field is similar for all QDs.

A key requirement for dragging is the presence of an unpaired
electron spin with a long spin-flip-time, either in the initial or
the final state of the optical transition; this condition is
satisfied by fundamental neutral ($\X$), single-electron ($\Xn$) and
single-hole ($\Xp$) charged QD transitions. The Overhauser field
\cite{Meier} experienced by this unpaired electron facilitates the
feedback that modifies the QD transition energy. However, whether or
not this feedback leads to \textit{resonance seeking} (as in the
blue Zeeman branch) or \textit{resonance avoiding} (as in the red
Zeeman branch) excitations depends on the spin orientation of the
electron that couples to the incident laser field.


\begin{figure*}[t]
\includegraphics[scale=0.84]{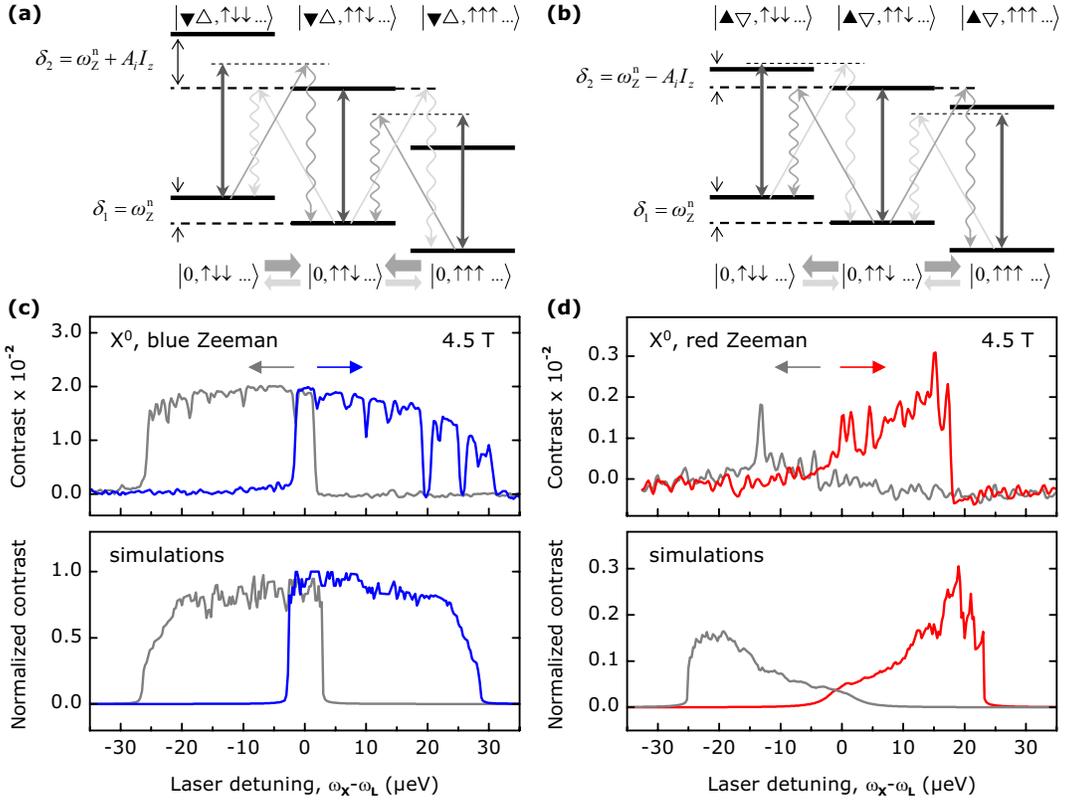}
\caption{Quantum dot nuclear spin polarization in resonant laser
scattering. Level diagrams of the blue (a) and red (b) Zeeman
transitions of a neutral exciton $\X$ in a finite magnetic field
applied along the growth direction $z$: resonant laser field
couples dipole allowed and dipole forbidden transitions (straight
and diagonal arrows, respectively) of the exciton-nuclear spin
manifold. For both Zeeman branches the lower states of the
manifold are electron/hole spin singlets $\ket{0}$ split by the
nuclear spin Zeeman energy
$\delta_1=\omega_{\mathrm{Z}}^{\mathrm{n}}$ according to their
nuclear spin orientation along $z$, e.g.
$\ket{\uparrow\downarrow\downarrow...}$ vs.
$\ket{\uparrow\uparrow\downarrow...}$. The upper states carry both
electron and hole spin excitations (full and open triangles,
respectively) and sense the nuclear field $I_{\mathrm{z}}$ of $N$
nuclei via the Overhauser shift $\pm (A/N)I_{\mathrm{z}}$ with the
hyperfine coupling constant $A/2$. Change in the nuclear spin
polarization occurs through spin-flip assisted diagonal
transitions followed by spin preserving radiative decay (wavy
arrows): finite laser detunings lead to an imbalanced competition
between the bidirectional nuclear spin diffusion processes within
the manifold (horizontal arrows). The coupled exciton-nuclear spin
system reaches steady state by locking the blue Zeeman transition
to the laser (dragging) or, alternatively, pushing the red Zeeman
transition away from the laser resonance (anti-dragging). The
corresponding spectra for opposite scan directions are color-coded
in (c) and (d) as grey and blue/red for initial red and blue laser
detunings (upper panel: experiments on sample A, lower panel:
simulations; note the factor of $\sim 10$ difference in ordinate
scales).} \label{fig-X}
\end{figure*}


We studied individual InGaAs QDs embedded in a field effect device
\cite{Drexler1994}. Two samples distinct by the thickness of the
tunnel barrier between the heavily $n$-doped back contact and the QD
layer (25~nm and 35~nm in samples A and B, respectively) were
employed to probe the fundamental exciton transitions in resonant
laser scattering experiments at 4.2~K~\cite{Hogele2004}.
Representative spectra measured on the neutral exciton $\X$ in
sample A subjected to a moderate magnetic field of $B_\m{z} = 4.5$~T
are shown in the upper panels of Fig.~\ref{fig-X}(c) and (d). The
spectra instantly reveal drastic departures from a two-level
Lorentzian with resonance frequency $\omega_{\mathrm{X}}$: the blue
Zeeman optical transition of $\X$ shows flat-top absorption
(Fig.~\ref{fig-X}(c)), also reported earlier for the negative trion
in Faraday \cite{Latta2009} and Voigt \cite{Xu2009} configurations.
In contrast, the spectral shape of the red Zeeman resonance is
triangular (Fig.~\ref{fig-X}(d)) with maximum contrast that is a
factor of $\sim 10$ lower than its blue counterpart. From direct
comparison it becomes apparent that the blue transition is locked to
the laser at frequency $\omega_{\mathrm{L}}$ and can be dragged to
positive and negative laser detunings
$\Delta=\omega_{\mathrm{X}}-\omega_{\mathrm{L}}$ by tens of $\mu$eV,
dependent on the scan direction (grey and blue spectra in
Fig.~\ref{fig-X}(c)). In contrast, the red transition avoids the
resonance with the laser using DNSP (Fig.~\ref{fig-X}(d)), resulting
in a triangular line shape.


\begin{figure}[t]
\includegraphics[scale=0.85]{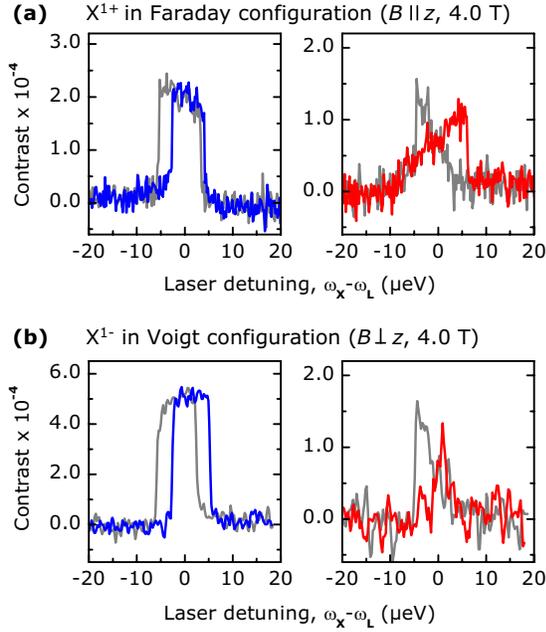}
\caption{Optical transitions in quantum dots of sample B at
$B=4.0$~T as a function of quantum dot charge and magnetic field
orientation: positive and negative trions, $\Xp$ and $\Xn$,
exhibit characteristic flat-top resonances on the blue Zeeman
transition and triangular lineshapes on the red Zeeman transition
(left and right panels in a and b, respectively) in magnetic field
oriented (a) parallel (Faraday) and (b) perpendicular (Voigt) to
the sample growth axis $z$.}\label{fig-XnXp}
\end{figure}


We systematically measure the same qualitative response for the blue
(red) Zeeman transitions of $\X$ as well as of both trions, $\Xp$
and $\Xn$, in Faraday and Voigt magnetic field geometries, as
demonstrated in Fig.~\ref{fig-XnXp}. There are, however,
quantitative variations in the efficiency of DNSP from sample to
sample and even from dot to dot within one sample~\cite{som}. The
thickness of the tunnel barriers in samples A and B plays a crucial
role for electron spin exchange with the Fermi reservoir via
co-tunneling~\cite{Smith2004} and thus for electron-spin pumping at
magnetic fields exceeding $\approx 0.3~\mathrm{T}$
\cite{Atature2006} as well as the efficiency of
DNSP~\cite{Latta2009}.

\begin{figure}[t,b]
\begin{center}
\includegraphics[scale=0.85]{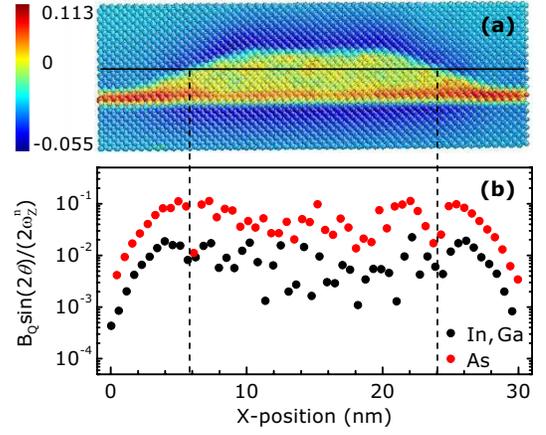}
\caption{(a) Distribution of the biaxial strain $\epsilon_\m{B}$,
which primarily controls the quadrupolar splitting, over the (100)
plane bisecting a truncated coned-shaped quantum dot. The false
color plot illustrates the tensile strain distribution with rapid
variations within the quantum dot region arising from the random
composition of the In$_{0.7}$Ga$_{0.3}$As alloy used in the model.
(b) The value of $B_\m{Q}^i \sin(2\theta_i)/(2\omega_\m{Z}^\m{n})$
evaluated over the line-cut through the monolayer indicated by the
solid line in (a). Averaging over this distribution with a
Gaussian envelope for the electron wavefunction we obtain a value
of 0.0062 for cations (In or Ga) and 0.0424 for anions
(As).}\label{fig-strain}
\end{center}
\end{figure}

Our findings demonstrate that effects of bidirectional DNSP are
omnipresent in resonant laser spectroscopy of QD excitons and call
for a unified explanation that goes beyond directional DNSP
mediated by flip-flop terms of the Fermi-contact hyperfine
interaction. Obviously, the model should be insensitive to the
details of the initial and final QD states, such as charge
configuration or the presence of dark exciton states
\cite{Bayer1999}, yet capture marked signatures and differences in
the response of the blue and red Zeeman transitions to a
near-resonant laser. Recently, Yang and Sham \cite{Yang2011}
proposed that non-collinear hyperfine interaction between
heavy-holes and the nuclei, induced by heavy-light-hole coupling,
provides an excellent qualitative description of the signatures
related to DNSP in resonant laser scattering \cite{footnote1}. On
the other hand, recent experiments \cite{Latta2011} demonstrate
that non-collinear hyperfine interaction  between the electron and
the nuclei plays a significant role in determining QD nuclear spin
dynamics even in the absence of optically generated holes; this
interaction is induced by large quadrupolar fields in strained
self-assembled QDs which ensure that nuclear spin projection along
$B_\m{z}$ is not a good quantum number. The resulting effective
non-collinear interaction between the QD electron and the nuclei
is~\cite{Huang2010}:

\begin{eqnarray}
    \hat{H}_\m{nc} &=& \sum_i A_i^\m{nc} \hat{I}_\m{x}^i \hat{S}_\m{z}
    \label{eqn:nc-hyperfine} \;.
\end{eqnarray}
with $A_i^\m{nc} = A_i B_\m{Q}^i
\sin(2\theta_i)/(2\omega_\m{Z}^\m{n})$, and $\hat{S}$,
$\hat{I}^i$, spin operators of the electron spin and the $i$-th
nucleus, respectively. Here, $\omega_\m{Z}^\m{n}$ denotes the
nuclear Zeeman energy, $B_\m{Q}^i$ the strength of the quadrupolar
interaction and $\theta_i$ is the angle between the major
quadrupolar axis of the $i$-th nucleus and the $z$-axis. For the
coupling strength of the electron to the $i$-th nucleus we assumed
$A_i = A/N$, where $A$ is the maximal Overhauser field splitting
and $N$ the number of nuclei. To determine $B_\m{Q}^i$ and
$\theta_i$, we first employed molecular statics with Tersoff type
force fields \cite{powell07} to obtain the realistic structure for
more than one million atoms hosting $N \simeq 32,000$ QD nuclei.
The atomistic strain and nuclear quadrupolar distributions are
extracted over this relaxed structure \cite{prb-indraft}.
Fig.~\ref{fig-strain}(a) shows the distribution of the biaxial
strain
$\epsilon_\m{B}\equiv\epsilon_\m{zz}-(\epsilon_\m{xx}+\epsilon_\m{yy})/2$
which is primarily responsible for the nuclear quadrupolar shifts.
Based on this distribution, we determine $A_i^\m{nc}$ for a
line-cut along the QD taken through the center and the $[010]$
axis, cf. Fig.~\ref{fig-strain}(b). Averaging over this
distribution for nuclei that lie within the Gaussian QD electron
wavefunction, we obtain $A_i^\m{nc} \simeq 1.3\cdot
10^{-4}~\mu$eV, consistent with~\cite{Latta2011}.

The fact that $\hat{H}_\m{nc}$ could explain dragging is at first
sight surprising since its dominant effect appears to be nuclear
spin diffusion. However, a careful inspection shows that the same
Hamiltonian also leads to a small polarization term whose
direction is determined by the sign of the optical
detuning~\cite{Yang2011,som}. To explain this, we consider the
energy level diagrams of $\X$ in Fig.~\ref{fig-X}(a) and (b), each
showing a ladder of two-level quantum systems coupled by nuclear
spin-flip processes. Here we adopt a mean-field description of the
nuclear spins by neglecting the quantum fluctuations in the
Overhauser field ($I_\m{z}=\langle\hat{I}_\m{z}\rangle$) and limit
ourselves to effective spin $1/2$ nuclei for simplicity. For a
given nuclear spin polarization $I_\m{z}$, e.g.
$\ket{\uparrow\uparrow\downarrow...}$, we can label the two level
system by the states $|0,I_\m{z}\rangle$ and
$\ket{\blacktriangledown\vartriangle , I_\m{z}}$. The ground
states $|0,I_\m{z}\rangle$ differ by an energy
$\delta_1=\omega_\m{Z}^\m{n}$. The corresponding energy
differences for the excited states are
$\delta_2=\omega_\m{Z}^\m{n}+A_iI_\m{z}$ and
$\omega_\m{Z}^\m{n}-A_iI_\m{z}$ for the blue and red Zeeman
transition, respectively. The transition rate associated with
hyperfine-assisted laser coupling is given for the blue Zeeman
branch (Fig.~\ref{fig-X}(a)) by
\begin{equation}
W_{\pm}(I_\m{z}) = \left(\frac{\Omega
A_i^\m{nc}}{4\omega_\m{Z}^\m{n}}\right)^2 \frac{\Gamma_0}{4
\delta_{\pm}^2 +\Gamma_0^2+\Omega^2/2},
\end{equation}
where $\Omega$ is the laser Rabi frequency and $\Gamma_0$ the
radiative decay rate \cite{footnote2}. A remarkable feature of
$W_{\pm}(I_\m{z})$ is its dependence on the sign of the laser
detuning entering through the effective optical detuning
$\delta_{\pm} = \Delta-A_i(I_\m{z}\pm 1) \mp\omega_\m{Z}^\m{n}$:
when the incident laser field is red (blue) detuned, the transition
rate $W_+(I_\m{z})$ ($W_-(I_\m{z})$) dominates over $W_-(I_\m{z})$
($W_+(I_\m{z})$) and ensures that the Overhauser field increases
(decreases). This directional DNSP will in turn result in a decrease
of the effective detuning $\delta$ from $\Delta - A_iI_\m{z}$ to
$\Delta - A_i(I_\m{z}+1)$ for a red detuned laser and to $\Delta -
A_i(I_\m{z}-1)$ for a blue detuned laser. If initially $I_\m{z} \ll
N/2$, then DNSP will continue until $\delta \simeq 0$.


\begin{figure}[t]
\includegraphics[scale=0.85]{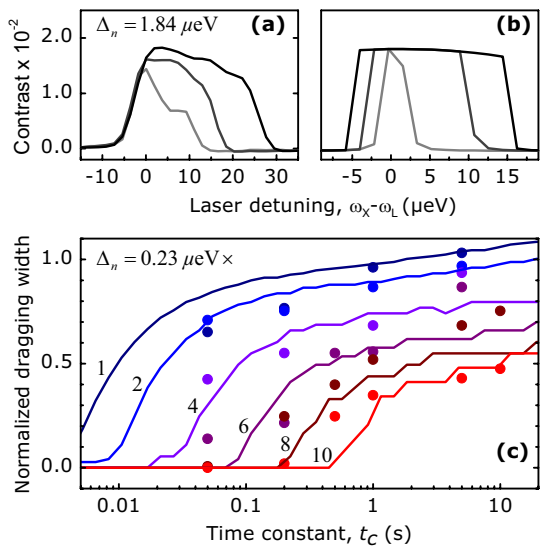}
\caption{Dragging width for the blue Zeeman branch of $\X$ in
sample A at 4.5~T as a function of discrete laser detuning step
$\Delta_n$ and integration time constant $t_\m{c}$. Experimental
spectra (a) and results of the simulations (b) obtained with steps
in laser detuning of $\Delta_n=1.84~\mu$eV at three time constants
$t_\m{c}$ of 0.2, 0.5 and 5.0~s (shown as grey, dark grey and
black solid lines, respectively). The parametric plot in (c)
depicts the dragging width obtained for incremental laser detuning
steps $\Delta_n=n\times 0.23~\mu$eV from $0.23~\mu$eV (top, dark
blue) to $2.30~\mu$eV (bottom, red) at different time constants
(closed circles: experiments, solid lines: simulations; the
parametric increment $n$ is given in numbers, the set of data was
normalized to the value of the dragging width for $n=1$ at 5~s).
The full dynamic range of the experiment is correctly reproduced
by the model.}\label{fig-speed}
\end{figure}


While a laser scan across the blue transition leads to a positive
feedback of the nuclear spins to ensure locking condition, a scan
across the red Zeeman line causes an anti-dragging effect. To
understand this, we note that the effective optical detuning in this
case is $\delta_{\pm} = \Delta  + A_i (I_\m{z} \pm 1)  \mp
\omega_\m{Z}^\m{n}$. The simple sign change in the effective optical
detuning renders the exact resonance between the laser field and the
red exciton transition an unstable point. The DNSP that ensues in
the presence of a small but non-zero $\delta$ will result in nuclear
spin flip processes that increase $|\delta|$ resulting in pushing
the red Zeeman transition away from the laser resonance. The
experiments validate these conclusions (Fig.~\ref{fig-X}(d) and
right panel of Fig.~\ref{fig-XnXp}).

To obtain a quantitative prediction, we consider the rate equation
\begin{equation}
\frac{dI_\m{z}}{dt} = W_{+}(I_\m{z}) (\frac{N}{2} - I_\m{z}) -
W_{-}(I_\m{z}) (\frac{N}{2}+I_\m{z}) - \Gamma_\m{d} I_\m{z}
\label{eqn:Iz}
\end{equation}
which includes nuclear-spin-flip assisted spontaneous emission
processes leading to pure nuclear spin diffusion at rate
$\Gamma_\m{d}$~\cite{som}. The steady-state solution exhibits
bistability due to the nonlinear $I_\m{z}$ dependence of the rates
$W_{\pm}(I_\m{z})$. For a given initial laser detuning $\Delta$ the
solution of the rate equation yields the steady state nuclear spin
polarization $I_\m{z}$ and thus the effective optical detuning
$\delta$ as established within the integration time $t_\m{c}$ of the
experiment. The absorption spectrum is calculated by varying the
laser detuning in discrete steps $\Delta_n$. The rate
equation~(\ref{eqn:Iz}) is symmetric with respect to the laser
detuning. In order to account for the asymmetry observed
experimentally for the two scan directions, we refined the model by
including terms for spin-flip Raman scattering processes that arise
from the Fermi-contact hyperfine interaction as well as an
unbalanced telegraph noise in the resonance condition~\cite{som}. We
find excellent agreement between theory and experiment, as
demonstrated in Fig.~\ref{fig-X}.

Remarkably, the model also reproduces the dependence of DNSP in
resonant laser scattering on key experimental parameters.
Fig.~\ref{fig-speed} shows how the dragging width evolves as a
function of scan speed. Both the detuning step $\Delta_n$ and the
waiting time constant $t_\m{c}$ used for signal integration after
each step contribute to the effective scan speed of the laser
detuning: for a given $\Delta_n$ the maximum width increases
non-linearly with $t_\m{c}$, as exemplified in
Fig.~\ref{fig-speed}(a) for $\Delta_n=1.84~\mu$eV. The
non-linearity in the functional dependence makes our simulations
highly sensitive to the set of parameters that determine the DNSP
dynamics; in particular, it allows us to extract a value for
$A_i^\m{nc}$ for a given set of $N$ and $A$.
Fig.~\ref{fig-speed}(c) demonstrates that the full dynamic range
of the experiment is correctly captured with the following set of
parameters: $\hbar \Gamma_0= 0.7~\mu$eV, $\Omega=0.5~\Gamma_0$,
$B=4.5$~T, step size $\Delta_n = 0.23~\mu$eV and dwell time
$t_\m{c}=0.2$~s, as used in the experiments, and
$\omega_\m{Z}^\m{e}/\omega_\m{Z}^\m{n}=1000$, $N = 3.2\cdot 10^4$,
$A_i= 240~\mu{\rm eV}/N$, $A_i^\m{nc}=0.45\cdot10^{-4}\mu$eV. The
value for the non-collinear hyperfine coupling constant found from
simulations is in good agreement with that obtained independently
from atomistic calculations and nuclear spin decay
measurements~\cite{Latta2011}. The same set of parameters was also
used to reproduce the external magnetic field and the laser power
dependence~\cite{som} and to calculate the absorption spectra in
Fig.~\ref{fig-X}.

Our results establish quadrupolar interaction induced non-collinear
hyperfine coupling as the mechanism responsible for resonant
bidirectional DNSP that is ubiquitous for self-assembled QDs. An
obvious extension of our work will be to carry out similar
experiments in interface or droplet QDs where quadrupolar
interactions are vanishingly small.


We acknowledge financial support from NCCR-Nanoscience, ERC, and the
DFG (SFB 631). We thank J. M. Sanchez, A. Badolato, D. Schuh and W.
Wegscheider for growing the two samples used in this work. We also
acknowledge useful discussions with M. Atature, N. Vamivakas, Y.
Zhang, M. Issler and P. Maletinsky.

\cleardoublepage

\setcounter{figure}{0} \setcounter{equation}{0}

\section{Supplementary online material}

{\bf Description of the experiment}

A voltage applied to the Schottky top gate was used to tune the QD
into stable charge configurations with the ground state being
singly positively charged, neutral or singly negatively
charged~\cite{Warburton2000-som,Seidl2005-som}. Single electrons
were injected into the QD from the Fermi reservoir, whereas in
absence of a hole reservoir in the device, single hole injection
was ensured by the presence of a weak non-resonant laser. The
sample was cooled in a bath cryostat to liquid helium temperature
(4.2~K) and finite magnetic fields were applied in Faraday or
Voigt configuration, parallel or perpendicular to the sample
growth axis $z$, respectively.

Resonant absorption of the neutral exciton $\X$ as well as
positive and negative trions, $\Xp$ and $\Xn$, was probed with a
tunable narrow-band laser in differential transmission
spectroscopy~\cite{Alen2003-som}. Absorption spectra were recorded
by setting the gate voltage or the laser energy to a specific
detuning $\Delta=\omega_{\m X}-\omega_{\m L}$, waiting for a time
$t_\m{c}$ and monitoring the transmission signal with a lock-in
amplifier. After each measurement, the voltage/laser energy was
changed by a discrete detuning step $\Delta_n$.

In finite magnetic fields, QDs in both samples A and B showed
pronounced dragging. QDs in sample A, for example, exhibited
dragging of both $\X$ and $\Xn$ in the range of tens of $\mu$eV
(Fig.~\ref{fig-X}) in contrast to QDs of sample B with sub 10
$\mu$eV scale for both trions (Fig.~\ref{fig-XnXp}). The thickness
of the tunnel barrier strongly affects electron spin exchange with
the Fermi reservoir via co-tunneling~\cite{Smith2004-som} and thus
the efficiency of DNSP~\cite{Latta2009-som}. In particular, in
sample B the dragging widths of $\Xn$ are reduced due to strong
spin pumping at magnetic fields exceeding $\approx 0.3~\mathrm{T}$
\cite{Atature2006-som}. In Voigt geometry the spin pumping leads
to a significant reduction in the transmission contrast and
consequently in the dragging efficiency even at the edge of the
charging plateau. However, the $\X$ of the same QD from sample B
exhibited a $\approx 20~\mu \mathrm{eV}$ dragging width,
consistent with the findings from sample A. For $\Xp$ the reduced
DNSP range is consistent with the presence of a non-resonant laser
which not only injects holes into the ground state of $\Xp$ but
also non-geminate electron spins opening up an additional nuclear
spin decay channel. The nuclear spin polarization is therefore
reduced for (photo-generated) single-hole-charged initial states.

{\bf Non-collinear hyperfine interaction}

The effective non-collinear hyperfine interaction Hamiltonian stems
from the fact that the quadrupolar interaction Hamiltonian for a
nuclear spin with strain axis tilted by an angle $\theta$ from the
$z$-axis (in the $x-z$ plane)
\begin{eqnarray}
    \hat{H}_\m{quad} &=& B_\m{Q} [\hat{I}_\m{z}^2 \cos^2\theta + (\hat{I}_\m{z}
    \hat{I}_\m{x} + \hat{I}_\m{x} \hat{I}_\m{z}) \sin\theta \cos\theta \nonumber \\
    &+&
    \hat{I}_\m{x}^2 \sin^2\theta] \;
    \label{eqn:quadrupolar}
\end{eqnarray}
does not commute with the dominant $\hat{H}_\m{fc,z} = \sum_i A_i
\hat{I}_\m{z}^i \hat{S}_\m{z}$ term of the Fermi-contact hyperfine
interaction $\hat{H}_\m{fc}$. To obtain an analytic expression for
$A_i^\m{nc}$, we assume $\theta \ll 1$ and use a Schrieffer-Wolff
(SW) transformation to obtain $\hat{H}_\m{hyp} = \hat{H}_\m{fc} +
\hat{H}_\m{hyp-quad}$ where
\begin{eqnarray}
    \hat{H}_\m{hyp-quad} &=& A_i^\m{nc}
    \hat{S}_\m{z} [\hat{I}_\m{x}^i \hat{I}_\m{z}^i + \hat{I}_\m{z}^i \hat{I}_\m{x}^i]
    \label{eqn:quadrupolar-hyperfine} \;,
\end{eqnarray}
with $A_i^\m{nc} = A_i B_\m{Q}^i
\sin(2\theta_i)/(2\omega_\m{Z}^\m{n})$. In $\hat{H}_\m{hyp-quad}$
we have only kept the terms that describe processes which leave
the electron spin-state unchanged, since contributions that flip
the electron spin will be negligible at high external fields.

Finally, we note that even for large $B_\m{z}$, the dominant role
of flip-flop terms of Fermi-contact hyperfine interaction is to
induce indirect interaction between the QD nuclei
\cite{Latta2011-som}: the primary effect of this interaction, in
the presence of fast optical dephasing of the electonic spin
resonance, is to ensure that the nuclear spin population assumes a
thermal distribution on timescales fast compared to the
polarization timescale determined by $A_i^\m{nc}$. In this limit,
the dynamics due to $\hat{H}_\m{hyp-quad}$ will be
indistinguishable from that described by $\hat{H}_\m{nc}$.

Averaging over the distribution for nuclei that lie within the
Gaussian QD electron wavefunction, we obtain for cations (In and
Ga with 9/2 and 3/2 nuclear spins, respectively, and
$A^\m{In}=112~\mu$eV, $A^\m{Ga}=84~\mu$eV \cite{Braun2006prb-som})
$A_i^\m{nc} \simeq 0.0062\cdot A^\m{In, Ga}$ and for anions (As
with 3/2 nuclear spin and $A^\m{As}=92~\mu$eV
\cite{Braun2006prb-som}) $A_i^\m{nc} \simeq 0.0424 \cdot
A_i^\m{As}$. For a fully polarized In$_{0.7}$Ga$_{0.3}$As system
we determine the value for the maximum Overhauser splitting due to
the non-collinear hyperfine coupling as
$A^\m{nc}=0.0062\cdot0.5\cdot(0.7\cdot
\frac{9}{2}A^\m{In}+0.3\cdot\frac{3}{2}A^\m{Ga})+0.0424\cdot0.5\cdot\frac{3}{2}A^\m{As}
\simeq 4.14~\mu$eV and obtain an average value of $A_i^\m{nc}
\simeq 1.3\cdot 10^{-4}~\mu$eV with $N=3.2\cdot10^{4}$.

Next, we discuss the derivation of the rate equation for the neutral
exciton blue Zeeman transition. We consider the limit of a large
external magnetic field where $\omega_\m{Z}^\m{e} \gg
\omega_\m{Z}^\m{n} \gg \Omega$, with $\Omega \sim \Gamma_0$
($\omega_\m{Z}^\m{e}$, $\omega_\m{Z}^\m{n}$ are the electron and
nuclear Zeeman energies, $\Omega$ the laser Rabi frequency and
$\Gamma_0$ the radiative decay rate). In this limit, all nuclear
spin-flip processes, including those described by $\hat{H}_\m{nc}$
are energetically forbidden to first order in perturbation theory.
Eliminating $\hat{H}_\m{nc}$ by a SW transformation we arrive at the
following correction terms to the laser-exciton coupling
\begin{equation}
\hat{H}_\m{nc-laser} = \m{i} \sum_i{\frac{\Omega
A_i^\m{nc}}{2\omega_\m{Z}^\m{n}}\left((\hat{\sigma}_\m{0X}-\hat{\sigma}_\m{X0})
\hat{I}^i_y\right)} \label{eqn:nc-laser}
\end{equation}
with $\hat{\sigma}_{0\m{X}}=\ket{0}\bra{\m{X}}$. Here,
$|\m{X}\rangle$ and $|0\rangle$ denote the exciton and vacuum state,
respectively. Application of the same SW transformation to the
Liouvillian term leads to nuclear-spin-flip assisted spontaneous
emission terms with maximum rate $\simeq \Gamma_0
(A_i^\m{nc}/4\omega_\m{Z}^\m{n})^2$. In the limit
$\omega_\m{Z}^\m{n} \sim \Gamma_0$ of interest, the denominator in
Eq.~(\ref{eqn:nc-laser}) should be modified to take into account
broadening of the excitonic spin states due to spontaneous emission.

The rates for spin-flip Raman scattering processes arising from
the Fermi-contact hyperfine interaction take place at a rate
$\simeq \Gamma_0 (A_i/4\omega_\m{Z}^\m{e})^2$; given that
$\omega_\m{Z}^\m{e} \simeq 1000~ \omega_\m{Z}^\m{n}$ and
$A_i^\m{nc} \simeq 0.02 A_i$, we conclude that the latter
processes will take place at a rate that is $\sim 300$ times
slower.

{\bf Modelling of the experimental data}


\begin{figure}[t]
\includegraphics[scale=0.85]{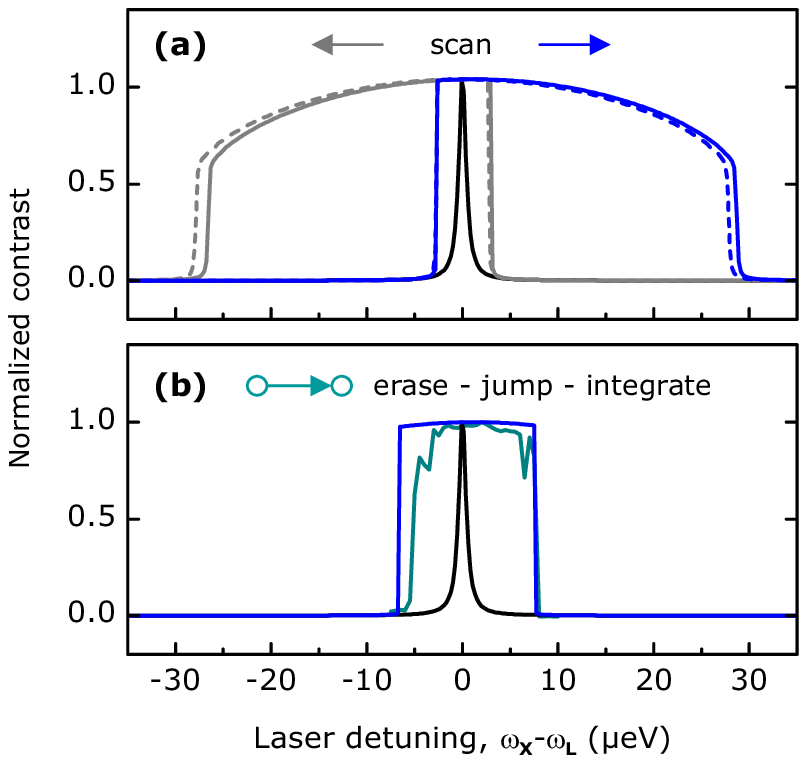}
\caption{(a) Simulations of dynamic dragging as in Fig.~1 of the
manuscript yet without spectral jitter: the solid and dashed lines
represent steady state solutions for scans ($t_\m{c}=0.2$~s and
$\Delta_n=0.23~\mu$eV) with and without spin-flip Raman scattering
processes according to Eqn.s~\ref{eqn:Iz-ROH-som} and
\ref{eqn:Iz-som}, respectively. The slight asymmetry to positive
laser detunings is a result of directional dynamic nuclear spin
polarization stemming from spin-flip Raman processes. (b)
Comparison between experimental spectra (dark cyan) and results of
the simulation (blue) for sequential data acquisition
($t_\m{c}=60$~s and $\Delta_n=0.46~\mu$eV): each data point of the
spectra was obtained by (i) erasing the nuclear spin polarization
in a voltage region with strong co-tunneling, (ii) subsequently
establishing a finite laser detuning, and (iii) integrating for
the time $t_c$ at this specific detuning. The Lorentzian spectra
in (a) and (b) are calculated with $\Gamma_0=0.73~\mu$eV and shown
in black for reference.}\label{fig-SOM-spectra}
\end{figure}



\begin{figure}[t]
\includegraphics[scale=0.85]{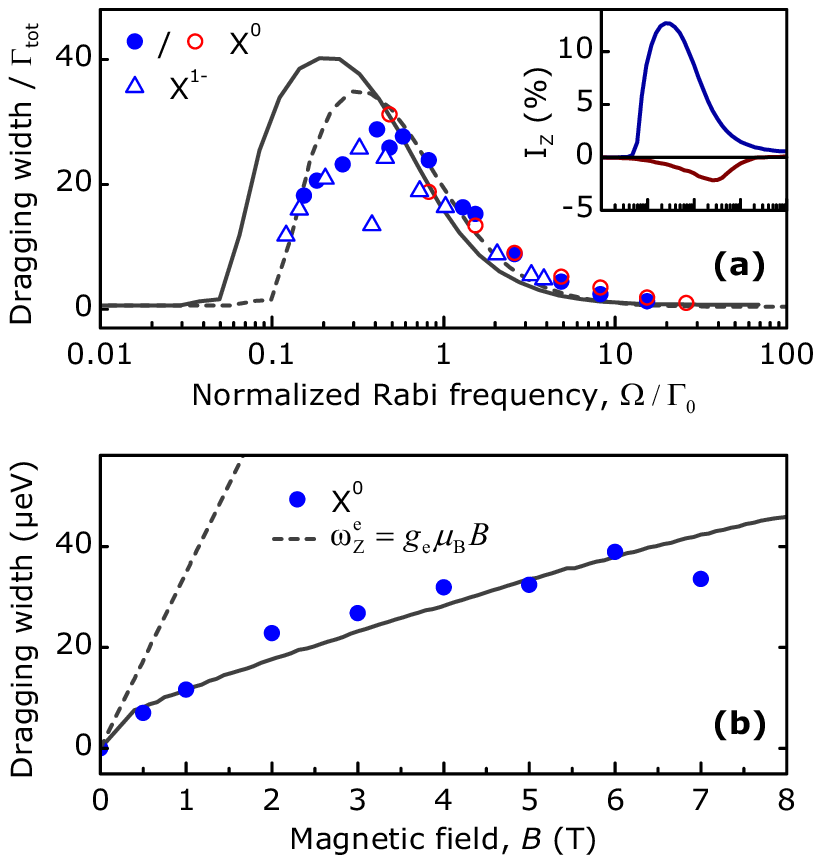}
\caption{(a) Laser power dependence of the maximum dragging width
in forward scans at $B=4.5$~T for the neutral exciton $\X$ (blue
Zeeman branch: closed circles; red Zeeman branch: open circles)
and the negative trion $\Xn$ (open triangles) in sample A. The
dragging width is normalized to the power-broadened total
linewidth $\Gamma_{\m{tot}}$, the laser power is expressed as
$\Omega/\Gamma_0$ (with experimentally determined radiative decay
rates $\Gamma_0=0.8$~ns$^{-1}$ for $\X$ and $1.1$~ns$^{-1}$ for
$\Xn$). The results of the model are shown by the solid line for
$\Omega/\Gamma_0$ and dashed line for $\Omega/(2\Gamma_0)$. The
inset shows on the same abscissa scale in red and blue the
corresponding degree of nuclear spin polarization accumulated in a
forward scan at initially negative and positive laser detunings,
respectively. Small but finite degree of nuclear spin polarization
at high laser powers stems from directional spin-flip Raman
scattering processes. (b) Maximum dragging width of $\X$ (circles)
as a function of magnetic field for $\Omega \simeq 0.5 \Gamma_0$.
The sub-linear monotonic increase of the maximally achieved
dragging width is reproduced by the simulations (solid line). The
linearly increasing electron Zeeman energy $\omega_{\m Z}^{\m e}$
is also shown (dashed line).}\label{fig-SOM-power}
\end{figure}


To obtain a quantitative prediction, we consider the rate equation
\begin{equation}
\frac{dI_\m{z}}{dt} = W_{+}(I_\m{z})(\frac{N}{2} + I_\m{z}) -
W_{-}(I_\m{z})(\frac{N}{2} - I_\m{z}) - \Gamma_\m{d} I_\m{z},
\label{eqn:Iz-som}
\end{equation}
where
\begin{equation}
\Gamma_\m{d} =\Gamma_0\left( \frac{A_i^\m{nc}}{4\omega_\m{Z}^\m{n}}
\right)^2\frac{\Omega^2/4}{\delta^2+\Gamma_0^2/4+\Omega^2/2}
\label{eqn:Gamma-d}
\end{equation}
\noindent is the rate at which nuclear-spin-flip assisted
spontaneous emission, leading to pure nuclear spin diffusion,
takes place. Here, as well as in the following equation,
$\delta=\Delta-A_i I_\m{z}$. The rate equation yields symmetric
dragging (dashed spectra in Fig.~\ref{fig-SOM-spectra}(a)) and
anti-dragging to either side of the resonance for the blue and red
Zeeman branches, respectively, qualitatively similar to the
results of Yang and Sham~\cite{Yang2011-som}.

Taking into account uni-directional spin-flip Raman scattering
processes that arise from the Fermi-contact hyperfine interaction,
we arrive at a refined rate equation model:
\begin{eqnarray}
\nonumber \frac{dI_\m{z}}{dt} &=&
W_{+}(I_\m{z})(\frac{N}{2}+I_\m{z}) -
W_{-}(I_\m{z})(\frac{N}{2}-I_\m{z}) - \Gamma_\m{d} I_\m{z} \\
&-& \Gamma_\m{sf}(\frac{N}{2} + I_\m{z}) \label{eqn:Iz-ROH-som}
\end{eqnarray}
with
\begin{equation}
\Gamma_\m{sf} =\Gamma_0\left( \frac{A_i}{4 \omega_\m{Z}^\m{e}}
\right)^2\frac{\Omega^2/4}{\delta^2+\Gamma_0^2/4+\Omega^2/2}.
\label{eqn:Gamma-d}
\end{equation}
Spin-flip Raman scattering processes at rate $\Gamma_\m{sf}$ give
rise to the asymmetry in the spectra for forward and reverse scan
directions (compare solid and dashed spectra in
Fig.~\ref{fig-SOM-spectra}(a)), in agreement with experimental
findings. The argument for the asymmetry holds when the nuclear
spin polarization is erased before a sudden jump to a finite
detuning and subsequent build-up of DNSP~\cite{Latta2009-som}:
locking of the resonance extends further for positive laser
detunings (Fig.~\ref{fig-SOM-spectra}(b)). Moreover, spin-flip
Raman scattering processes ensure small but finite nuclear spin
polarization at high laser powers, as shown in the inset of
Fig.~\ref{fig-SOM-power}(a).

The normalized absorption spectra depicted in
Fig.~\ref{fig-SOM-spectra}(a) are calculated from steady state
solutions of Eqn.s~\ref{eqn:Iz-som} and \ref{eqn:Iz-ROH-som} with
the following parameters: $\hbar \Gamma_0= 0.7~\mu$eV from the
radiative decay rate $1/\Gamma_0 = 1.2$~ns determined from
saturation~\cite{Hogele2004-som}, $\Omega=0.5~\Gamma_0$,
$B=4.5$~T, step size $\Delta_n = 0.23~\mu$eV and dwell time
$t_\m{c}=0.2$~s, as used in the experiments, and
$\omega_\m{Z}^\m{e}/\omega_\m{Z}^\m{n}=1000$, $N = 3.2\cdot 10^4$,
$A_i= 240~\mu{\rm eV}/N$, $A_i^\m{nc}=0.45\cdot10^{-4}~\mu$eV.
Intrinsic decay of the nuclear spin polarization is negligible for
the ground state of $\X$~\cite{Maletinsky2009-som,Latta2011-som}.
Here, we omitted the unbalanced telegraph noise in the resonance
condition used to calculate the spectra in Fig.~1 of the
manuscript. This jitter in the resonance condition with an
amplitude of 0.5 $\mu$eV (smaller than the linewidth) and
timescales longer than $t_\m{c}$ was included in simulations in
order to account for the asymmetry observed experimentally for the
two scan directions. It is consistent with the spectral
fluctuations in resonant QD spectroscopy~\cite{Hogele2004-som}.
Based on experimental observations for the QDs in sample A, it is
reasonable to assume these fluctuations to be unbalanced with a
small weight on the higher energy side of the resonance. However,
comparing the observed traces for different QDs in different
samples reveals that the spectral jitter can actually appear on
either side of the resonance.

The dependence of the dragging width on the magnitude of the laser
power and the external magnetic field provides further
confirmation of the model. Fig.~\ref{fig-SOM-power} shows how the
dragging width evolves as a function of laser power and magnetic
field. The effect of dragging is inhibited at low incident powers,
increases until reaching a maximum below the saturation at $\Omega
\simeq \Gamma_0$, and vanishes in the limit of high excitation
powers (Fig.~\ref{fig-SOM-power}(a)). This is consistent with the
prediction of Eq.s~\ref{eqn:Iz-som} and \ref{eqn:Iz-ROH-som}: the
maximum dragging width is expected at $\Omega<\Gamma_0$ for
non-vanishing $\Gamma_{\m d}$ and $\Gamma_{\m{sf}}$. Both $\X$ and
$\Xn$ reveal the same dependence when the dragging width on the
ordinate is normalized to the total linewidth
$\Gamma_{\mathrm{tot}}=\sqrt{\Gamma_0^2+\Omega^2/2}$ and the
abscissa is expressed is units of $\Omega$/$\Gamma_0$. The results
of the model reproduce our experimental findings (solid line in
Fig.~\ref{fig-SOM-power}(a)) predicting $\sim 10~\%$ of nuclear
spin polarization at maximum (inset of
Fig.~\ref{fig-SOM-power}(a)). The model overestimates the dragging
width at powers below saturation but gives perfect agreement above
saturation for $\tilde{\Omega}=\Omega/2$. We speculate that the
scaling factor of 2 stems from the line broadening $\Gamma \simeq
2~\Gamma_0$ that we typically find in our samples as a result of
spectral fluctuations~\cite{Hogele2004-som}. The monotonic
sublinear increase of the dragging width with magnetic field, as
measured on $\X$ close to saturation, is clearly reproduced by our
model (solid line in Fig.~\ref{fig-SOM-power}(b)) and provides
further confirmation for the quantitative nature of our analysis.


\begin{thebibliography}{}
\bibitem{footnote1} Experimentally determined values of hole-hyperfine interaction would suggest that the resulting nuclear spin-flip
rates would be an order of magnitude too small to explain the
relative insignificance of the directional reverse Overhauser
effect. In addition, the large apparent variation in
heavy-light-hole mixing in QDs as indicated by the measured
in-plane hole $g$-factors would suggest that dragging features
would change appreciably from QD to QD, if the mechanism was due
to hole-hyperfine interaction.
\bibitem{footnote2} By carrying out a complete numerical
analysis, we have confirmed that the saturation of
$W_{\pm}(I_\m{z})$ is correctly taken into account in Eq.~(2).
\end{thebibliography}
\end{document}